%% file: sn-article.tex
\documentclass[pdflatex,sn-mathphys-num]{sn-jnl}

\usepackage{graphicx}%
\usepackage{multirow}%
\usepackage{amsmath,amssymb,amsfonts}%
\usepackage{amsthm}%
\usepackage{mathrsfs}%
\usepackage[title]{appendix}%
\usepackage{xcolor}%
\usepackage{textcomp}%
\usepackage{manyfoot}%
\usepackage{booktabs}%
\usepackage{algorithm}%
\usepackage{algorithmicx}%
\usepackage{algpseudocode}%
\usepackage{listings}%
\usepackage{tabularx}
\usepackage{booktabs}
\usepackage{pifont}
\usepackage{svg}

\usepackage[textsize=tiny,
	]{todonotes}
\usepackage[normalem]{ulem}

\theoremstyle{thmstyleone}%
\theoremstyle{thmstyletwo}%

\theoremstyle{thmstylethree}%

\begin{document}

\title[RAGe: RAG Evaluation Framework]{RAGe: A Retrieval-Augmented Generation Evaluation Framework}


\author*[1]{\fnm{Larissa} \sur{Guder}}\email{larissa.guder@edu.pucrs.br}
\author[1]{\fnm{João Pedro} \sur{de Moura Medeiros}}\email{j.moura001@edu.pucrs.br}
\author[1]{\fnm{Arthur} \sur{Accorsi}}\email{arthur.accorsi@edu.pucrs.br}
\author[1]{\fnm{Gustavo} \sur{Losch do Amaral}}\email{g.losch@edu.pucrs.br}
\author[1]{\fnm{Maurício} \sur{Cecílio Magnaguagno}}\email{mauricio.magnaguagno@acad.pucrs.br}
\author[1]{\fnm{Felipe} \sur{Meneguzzi}}\email{felipe.meneguzzi@pucrs.br}
\author[1]{\fnm{Marcio Sorraglia} \sur{Pinho}}\email{marcio.pinho@pucrs.br}
\author[1]{\fnm{Dalvan} \sur{Griebler}}\email{dalvan.griebler@pucrs.br}

\affil*[1]{\orgdiv{School of Technology}, \orgname{Pontifical Catholic University of Rio Grande do Sul}, \orgaddress{\street{Av. Bento Gonçalves}, \city{Porto Alegre}, \state{Rio Grande do Sul}, \country{Brazil}}}


\include{sections/00_abstract}



\maketitle
\include{sections/01_intro}
\include{sections/02_background}

\include{sections/03_methodology}
\include{sections/04_toolchain}
\include{sections/05_results}
\include{sections/06_rw}
\include{sections/07_conclusion}

\backmatter

\bmhead{Acknowledgements}

This work was partially supported by DELL Technologies, FAPERGS 09/2023 PqG (N\textsuperscript{o} 24/2551-0001400-4), and CNPq Research Program (N\textsuperscript{o}311012/2025-6).

\begin{appendices}






\end{appendices}


\bibliography{rage}

\end{document}

%% file: sections/00_abstract.tex
\abstract{
Deploying Large Language Model (LLM) applications, particularly those relying on Retrieval-Augmented Generation (RAG), remains challenging due to high computational demands, outdated knowledge bases, and the need to manually select optimal pipeline components. 
In this work, we propose a modular framework for benchmarking and guiding the efficient development of RAG applications by focusing on resource telemetry and component recommendation, suggesting the best components for a domain-specific dataset. 
Our approach leverages core techniques in LLM applications, including document chunking, vector databases, embedding models, and retrievers, to evaluate trade-offs among accuracy, efficiency, and scalability. 
By directly correlating retrieval and generation quality with underlying hardware constraints, RAGe supports researchers to identify the most effective, domain-specific RAG setups for their specific operational needs, facilitating rapid prototyping even on consumer-grade hardware.


}
\keywords{Natural Language Processing, Retrieval-Augmented Generation, Large Language Models, Text Embedding}

%% file: sections/01_intro.tex
 
\section{Introduction}\label{sec:intro}

Artificial Intelligence (AI) has become a crucial tool for technological advancement, driving innovation in industries such as healthcare, finance, education, and software development.
One of the most significant advances in this field has been the development of Large Language Models (LLMs), which show remarkable capabilities in understanding and generating text of human-level quality.
The creation and capabilities of these models gained public attention when OpenAI released GPT-3. 
Since then, numerous LLMs have been developed to support a wide range of applications in different domains, benefiting both individuals and organizations~\cite{10.3389/fcomp.2025.1523699}.

These models operate as black boxes, lacking mechanisms to verify the factual correctness of their outputs, which can lead to inaccurate information. 
Furthermore, as the demand for efficient AI solutions continues to increase, addressing these limitations has become a research problem with direct implications for industries and users that rely on LLM applications~\cite{kostikova2025lllms}.

As training data is fixed at the time of model creation, models cannot access up-to-date or domain-specific knowledge.
One approach to mitigate these challenges is Retrieval-Augmented Generation (RAG)[REF]. 
RAG integrates a retrieval component that allows the model to access external knowledge sources before generating a response. 
This process enriches the model input with up-to-date, domain-specific data, thereby reducing hallucinations and enhancing the accuracy and reliability of the generated output. 

Deploying LLM-based applications, particularly those relying on RAG, is a significant challenge. 
This includes high computational demands and the careful and rigorous selection of components, such as embedding models, reranking strategies, and similarity metrics. 
Most existing work focuses on large-scale architectures, ignoring lightweight alternatives. This approach requires professionals to manually evaluate each part of the pipeline to achieve satisfactory results, which is often time-consuming and resource-intensive.

Existing RAG-based frameworks in the literature, including ModularRAG~\cite{modular_rag} and FlashRAG~\cite{flashrag}, focus on providing a modular architecture for building RAG applications. 
However, after evaluating these frameworks, we observed that they lack a dedicated optimization module, which we identify as a key component to improve the performance and efficiency of RAG pipelines.

To fill this gap, we introduce RAGe, a novel evaluation framework designed for resource-constrained environments. Our contributions are threefold. 
First, we propose a standardized methodology and toolchain for systematically assessing small, open-source models and retrievers on consumer-grade hardware. 
Second, we evaluate the efficiency-accuracy trade-offs inherent to these pipelines, providing a comprehensive and reproducible baseline for local deployments. 
Third, we demonstrate that by isolating the impact of hyperparameter tuning and architectural configurations on multidimensional metrics, such as latency, memory consumption, and generation quality, RAGe enhances researchers' ability to discover optimal, scalable model retrievers tailored to specific hardware constraints and application contexts.

This paper is structured as follows: Section~\ref{sec:background} presents the main concepts that are used to build our framework, such as Large Language Models (LLMs), Retrieval-Augmented Generation (RAG), datasets, chunking, embeddings, and metrics.
Section~\ref{sec:methodology} presents the methodology, with a description of each component in our framework, how they are used and evaluated.
Section~\ref{sec:toolchain} presents an implementation of the previously presented framework as a toolchain.
Section~\ref{sec:results} presents the results of our approach, focusing on the output from the user interface and how it presents multiple result sessions and impacts user choice.
Section~\ref{sec:rw} presents the related work regarding RAG frameworks and a discussion about our findings, and ending with our conclusions in Section~\ref{sec:conclusion}.

%% file: sections/02_background.tex
\section{Background}\label{sec:background}
To provide the necessary context for the present article, this section reviews the fundamental concepts related to LLMs, RAG, datasets, chunking strategies, embeddings, retrieval, hybrid search, reranking, evaluation metrics, and storage architectures. 
The goal is to establish a theoretical foundation that supports the subsequent analysis and experiments, ensuring that each component of the system is understood within the AI applications frameworks.

\subsection{Large Language Models (LLMs)}\label{subsec:llm}

Large Language Models (LLMs) are deep learning models based on the Transformer architecture, introduced by Google~\cite{10.5555/3295222.3295349}. 
The development of LLMs began with models such as BERT~\cite{devlin-etal-2019-bert} and the GPT series~\cite{gpt}. 
These models have shown strong performance in Natural Language Understanding and Natural Language Generation,
setting a new standard for Natural Language Processing (NLP) systems.

The Transformer architecture uses self-attention mechanisms, enabling these models to capture long-range dependencies in text more effectively than previous recurrent neural network-based architectures~\cite{10.5555/3295222.3295349}. 
The self-attention mechanism allows the model to process all input tokens in parallel. 
This fundamental shift enabled massive computational scaling, making it feasible to train models with billions of parameters on massive text corpora.
With these mechanisms, it was possible to train Large Language Models (LLMs).

Each LLM contains a large number of trainable parameters, also known as weights.
For LLMs, the number of parameters influences the model's capacity to understand and perform tasks.
For example, GPT-3 has 175 billion parameters ~\cite{gpt}, and Llama has 65 billion parameters \cite{touvron2023llamaopenefficientfoundation}, allowing them to handle a wide range of language-related tasks.
In practice, LLMs have been successfully applied to a wide variety of tasks, including conversational agents, machine translation, text summarization, and sentiment analysis \cite{matarazzo2025survey}.
However, to improve performance and ensure output accuracy, other factors, such as temperature, top-k, and additional model parameters, must be tuned.
<<Aqui ocorre uma mudança brusca de assunto.>>
The adoption of chatbots has changed how humans communicate with machines. 
Through a user query, the LLM can understand what the user wants and generate an appropriate answer. 
In our work, the LLM is a critical component that processes user input and retrieves contextual data to generate relevant and coherent responses.

One of the key limitations of LLMs is that their knowledge is limited to the corpora used during the training stage, making it difficult to adopt these models in specific domains.
To address this, two approaches are commonly used: fine-tuning and RAG. 
Due to the high cost of fine-tuning on these large models, RAG is the most common method to overcome this limitation. 

\subsection{Retrieval-Augmented Generation (RAG)}\label{subsec:rag}
With the adoption of LLMs across companies and universities, it is necessary to adapt them to specific domains based on their intended use. 
Since LLMs are trained on static datasets, they lack access to information beyond their knowledge cutoff date. This makes them prone to generating outdated or factually erroneous information and to hallucinating.

In domains such as politics or economics, models must be continuously updated with current news and financial developments. 
This is important for investors and companies in the area, as they must stay informed about current financial matters~\cite{AssessingLargeLanguageModels}. Other research areas show the same problem, as in Medicine, where models must adapt to new clinical evidence and medical research, to ensure the accuracy and safety of their predictions \cite{wu2025assessing}.

Another important problem here is performance degradation when models are trained on small or outdated datasets, which can lead to overfitting, poor generalization, and incorrect predictions. 
However, periodically retraining models is computationally expensive, so there is a trade-off between staying current and the costs of updating \cite{ibrahim2024simple}.

To address these problems, Retrieval-Augmented Generation (RAG) was proposed. 
It involves retrieving relevant information from external knowledge bases, including documents, metadata, databases, and other sources, answering user questions based on the retrieved information, and complementing the prompt with that information for LLMs~\cite{10.1145/3637528.3671470}. 
It is a method that provides context for the LLM by using a user's query about a specific segment to enrich the prompt on the server side. 
RAG is a technique similar to fine-tuning, but it does not require retraining the entire model to learn from the data domain and update its weights, as it provides the context of the document by prompt engineering~\cite{10.5555/3495724.3496517}.

\subsection{Datasets}\label{subsec:dt}

Evaluating a RAG framework requires ground-truth-annotated datasets. 
Then it is possible to determine if the chosen configurations meet the user's needs. 
For RAGe, we selected the following datasets: NaturalQuestions (Wikipedia)~\cite{kwiatkowski-etal-2019-natural}, NewsQA (CNN news)~\cite{trischler-etal-2017-newsqa}, and TriviaQA (general knowledge)~\cite{joshi-etal-2017-triviaqa}. 
These datasets were selected from the literature and are widely used as benchmarks for testing and comparison.

We define these datasets for evaluation because they include ground-truth answers, which can be used as a validation set. 
Each dataset contains over 100,000 instances, allowing for the evaluation of various models and configurations. 
It is important to highlight that in RAGe a user might include their own datasets from other domains.
This is important because of the domain where the application will be used, with a dataset directly related to the main problem of the final user, the RAG configuration can be more consistently evaluated. 

\begin{table}[ht]
    \caption{Structural information of QA datasets.}
    \label{tab:qa_datasets}
    \centering
    \begin{tabular}{|l|c|c|c|}
        \hline 
            \textbf{Dataset} & 
            \textbf{Natural Questions} & 
            \textbf{NewsQA} & 
            \textbf{TriviaQA} \\
            \hline
                Source &
                Wikipedia (Google Search) & 
                CNN News Articles & 
                Trivia / Web / Wikipedia \\
            \hline
                Size &
                $\sim$328 mb &
                $\sim$358 mb &
                $\sim$604 mb \\
            \hline
                Train (size) & 
                307,373 examples &
                380,000 examples &
                650,000 examples \\
            \hline
                Question Type & 
                Real user queries & 
                Human-written & 
                General knowledge \\
            \hline
            Access link & 
            \shortstack[l]{ai.google.com/research/\\NaturalQuestions/dataset} &
            \shortstack[l]{cs.nyu.edu/\\$\sim$kcho/DMQA/} &
            \shortstack[l]{nlp.cs.washington.edu/\\triviaqa/\#data}\\
        \hline
    \end{tabular}

\end{table}

\subsection{Chunking}
Chunking is a crucial step in NLP that involves breaking documents, paragraphs, sentences, or text into smaller pieces, known as chunks. 
This step is essential because it preserves semantic coherence, enabling the system to achieve more accurate and efficient performance, improving tasks such as embedding. 
Different approaches can be applied for chunking, including fixed-size chunks, recursive chunks, and semantic chunks. 

In RAGe, this task is handled by a module called the splitter, which processes the input text and divides it into chunks according to the user's choice. Each resulting chunk is then passed to the embedding module to generate the representations. 

\subsection{Embeddings}\label{subsec:embeddings}
Embedding is a technique widely used in LLMs and various AI applications to represent unstructured data, such as images, videos, or text, in a vector format \cite{jing2402large}. 
In text, there are different levels of granularity, such as sub-words, words, sentences, or documents. The embedding model will convert text into dense numerical vectors, typically with fixed dimensions such as 768 or 1024~\cite{nastase2023grammatical}.

\begin{figure}[H]
    \centering
    \includegraphics[width=0.8\linewidth]{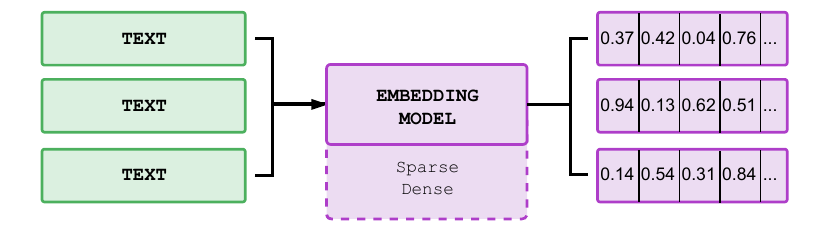}
    \caption{Overview of the embedding process applied in RAGe. The input texts are processed by an embedding model, which generates vector representations that encode the semantic content of the original text.}
    \label{fig:embedding-process}
    
\end{figure}

In our case, we focus on sentence embeddings, where sentences are represented as vectors keeping context, with related terms positioned closer together in the vector space. 
The process involves creating a dense, high-dimensional vector that is similar to the data, capturing its syntactic and semantic meaning. Figure~\ref{fig:embedding-process} shows how the embedding process is applied to RAGe.

This method enables comparison between the embedding vector of a user query and the embedding vectors of the documents, allowing for the retrieval of the most semantically relevant results.
In RAGe, two models of embedding "all-MiniLM-L6-v2"~\footnote{Embedding model available at https://huggingface.co/sentence-transformers/all-MiniLM-L6-v2} and "jina-embeddings-v3"~\footnote{Embedding model available at https://huggingface.co/jinaai/jina-embeddings-v3} are included in the architecture.

\subsection{Retrieval Strategies}\label{subsec:retrieval}
Based on the information retrieved by RAG, the LLM will generate an answer to the user's prompt that is relevant to that information. 
The objective is to ensure that the most similar chunks are obtained. 
Two main retrieval strategies are considered, including state-of-the-art techniques such as Hybrid Search and Reranking, enhancing the quality of the retrieved results. 

In addition, our architecture supports more traditional approaches, such as Similarity Search (also known as semantic search) and keyword-based retrieval or lexical search. 
Some retrieval techniques, such as semantic chunking, have been discarded; the literature states that the results are similar, but they come with higher computational costs~\cite{qu-etal-2025-semantic}.

\subsubsection{Hybrid Search}\label{subsubsec:hs}
Hybrid Search is a retrieval technique~\cite{hybridSearch} that combines lexical and vector-based searches to answer a user query more effectively, integrating sparse retrieval models (e.g., BM25~\cite{SPARCKJONES1972}).
Lexical search focuses on matching exact words from the query to those of the documents \cite{gao2021coil}.
While vector search transforms the user query into an embedding and compares it with documents, it also applies an embedding to capture semantic similarity \cite{semanticsimilarity}.
By integrating these two approaches, hybrid search provides robustness and result quality for the user. 
In RAGe, hybrid search is implemented as an optional component.

\subsubsection{Reranking}\label{subsubsec:reranking}
Reranking refers to the process of ordering a set of documents retrieved from a hybrid or lexical search based on their semantic similarity to the user query. 
In RAG systems, this step often uses embedding similarity to sort documents, which enhances the relevance of the final retrieved context used by the LLM~\cite{nogueira2020passagererankingbert}. RAGe uses jina reranker v2 as its reranking model~\footnote{Reranking model available at https://huggingface.co/jinaai/jina-reranker-v2-base-multilingual}.

\subsection{Metrics}\label{subsec:metrics}
While traditional evaluation metrics remain relevant, the increasing complexity of language models has led to the development of more specialized metrics. 
In the context of LLMs, specific metrics such as hallucination, faithfulness, and relevance of the answers are applied to evaluate how accurately a model generates responses based on retrieved knowledge.

In addition, there are similarity metrics to measure the distance between embedding vectors. 
In RAGe, metrics such as the euclidean distance and cosine similarity were used to evaluate the effectiveness of embedding-based components.

\subsubsection{Hallucination}
In the context of LLMs, hallucination\footnote{LLM metric available at https://deepeval.com/docs/metrics-hallucination} refers to models that give facts or generate responses that do not exist and do not make sense in the context provided. 
The metric uses an LLM to determine, for each context in contexts, whether there are any contradictions to the ground-truth output~\cite{10.1145/3703155}. This issue often comes from the statistical nature of their knowledge.

\begin{equation}
    \label{eq:hallucination}
    \text{Hallucination} = 1 - 
        \frac{\text{Number of Contradicted Contexts}}
        {\text{Total Number of Contexts}}
\end{equation}
\text{\\}

\subsubsection{Faithfulness}
Faithfulness evaluates whether the response generated by the model remains true to the given context, ensuring that the responses are consistent with the information from the context~\cite{ragas}.

\begin{equation}
    \text{Faithfulness} =
        \frac{\text{Number of claims in the response supported by the retrieved context}}
        {\text{Total number of claims in the response}}
\end{equation}
\text{\\}

\subsubsection{Answer Relevancy} Answer Relevancy evaluates whether the answer generated by a model is suitable for the given question~\cite{ragas}.
To estimate it, for the given answer, the LLM generates \textit{n} potential questions. 
For each potential question, the similarity to the original question is calculated.

\begin{equation}
    \text{AR} =
        \frac{1}{n}
        \sum_{i=1}^{n}
            \text{cosine similarity}
                (\text{Given Answer},
                \text{ Potential Questions})
\end{equation}
\text{\\}

\subsubsection{Context Precision} The context precision\footnote{LLM metric available at https://deepeval.com/docs/metrics-contextual-precision} metric  uses an LLM to assess whether each node in the retrieved context is relevant to the input, based on the information provided in the expected output.

In the formula, each retrieved node is evaluated individually. 
The term k represents the position of each node within the list of retrieved nodes, ordered from 1 to n. 
The term n refers to the number of elements retrieved as context. 
For each node, a binary indicator value of 1 indicates that the node is considered relevant to the input and expected output, while a value of 0 indicates otherwise.
Finally, the cumulative value is used to calculate the average.

\[rk = 
\begin{cases}
    1, & \text{for nodes that are relevant}\\
    0, & \text{for nodes that are not relevant}
\end{cases}
\]

\begin{equation}
    \text{CP} = 
        \frac{\sum_{k=1}^{n} 
            \left
                (\frac{\text{Amount of Relevant Nodes Up to Position } k}{k} \times rk
            \right)}{\text{Amount of Relevant Nodes}}
\end{equation}

\subsubsection{Context Recall}
The Context Recall\footnote{LLM metric available at https://deepeval.com/docs/metrics-contextual-recall} metric evaluates how comprehensively the retrieved context supports the expected output.

\begin{equation}
    \text{CR} = 
        \frac{\text{Number of Attributable Statements}}{\text{Total Number of Statements}}
\end{equation}
\text{\\}

\subsubsection{Euclidean Distance}\label{subsubsec:euclidean}

The Euclidean distance is the distance between two embedding vectors $(x,y)$. 
In vector databases, this distance is often used to determine the similarity between two textual representations, with smaller distances indicating higher semantic similarity.

\begin{equation}
    d(\mathbf{x}, \mathbf{y}) = 
        \sqrt{\sum_{i=1}^{n} (x_i - y_i)^2}
\end{equation}

\subsubsection{Cosine Similarity}\label{subsubsec:cosine}

The Cosine similarity is the angle between two vectors $(x,y)$, indicating how similar they are in direction.
In vector databases, cosine similarity is often chosen over Euclidean distance, as it focuses on the direction of the vectors rather than their magnitude, making it more robust when embeddings have different scales.

\begin{equation}
    \cos(\theta) = 
        \frac{\mathbf{x} \cdot \mathbf{y}}{\lVert \mathbf{x} \rVert \lVert \mathbf{y} \rVert}
\end{equation}

\subsection{Storage Types}\label{subsec:storage_type}

Once the information is converted into vectors (Figure~\ref{fig:embedding-process}), it must be stored in a storage that enables fast and accurate retrieval, especially when dealing with large volumes of data. This avoids the need to process the entire dataset for each execution, ensuring efficiency and scalability in queries. There are two approaches to storage types: vector databases and vector libraries.

\subsubsection{Vector Databases (VDBs)}\label{subsec:vdb}
Vector databases (VDBs) are designed to handle and store vector representations for computing distance metrics, as seen in Subsections~\ref{subsubsec:euclidean} and~\ref{subsubsec:cosine}, allowing for efficient comparisons of similarity. 
These databases are commonly used in applications that require high speed and accuracy \cite{pan2024survey}.

In RAGe, the VDBs play a central role, serving as the core infrastructure for handling all vectorized data.
Using similarity metrics, it will rank the most similar embeddings with the user input, which is first converted into an embedding vector.
In RAGe, we integrate Milvus and PGvector as VDB solutions.

Milvus\footnote{Vector database available at https://milvus.io} is a vector database designed for high-performance similarity search and retrieval of large-scale vector data.
It supports a variety of indexing and search algorithms, making it suitable for scalable applications.

PGVector\footnote{Vector database available at https://github.com/pgvector/pgvector}is a PostgreSQL extension that adds support for vector data types and similarity search within the PostgreSQL ecosystem. It allows vector storage and comparison using SQL queries, and avoids the operational overhead and complexity of deploying and maintaining new systems.

\subsubsection{Vector Libraries (VLs)}
Vector Libraries (VLs) have shown significant advantages over VDBs, mainly in speed and retrieval quality.
These systems are characterized by high performance, which is achieved by targeting specific workloads and query types. This specialization, however, results in a comparatively limited range of functional capabilities~\cite{pan2024survey}.
For example, it is not possible to add or remove a chunk without recreating the index.

VLs require careful configuration and parameter tuning to balance performance and ensure data is maintained up to date. VLs provide code-based tools for similarity search, as well as operations and manipulation of high-dimensional vectors.
In RAGe, we adopted the FAISS library to enable scalable, high-performance vector retrieval.
FAISS~\cite{douze2024faiss} is a vector library that provides support for similarity search, evaluation, and parameter tuning. 
It is written in C++ and offers a Python wrapper, making it easy to use in simple scripts.

%% file: sections/03_methodology.tex
\section{The RAGe Framework}\label{sec:methodology}

Implementing a RAG framework requires substantial time and engineering effort. 
RAGe addresses this challenge by providing a modular architecture that allows users to effortlessly choose components, configure them, and prototype different RAG pipelines. 
Based on the selected parameters, RAGe automatically recommends an optimized configuration for a given scenario. 

\begin{figure}[ht]
    \centering
    \includegraphics[width=1\linewidth]{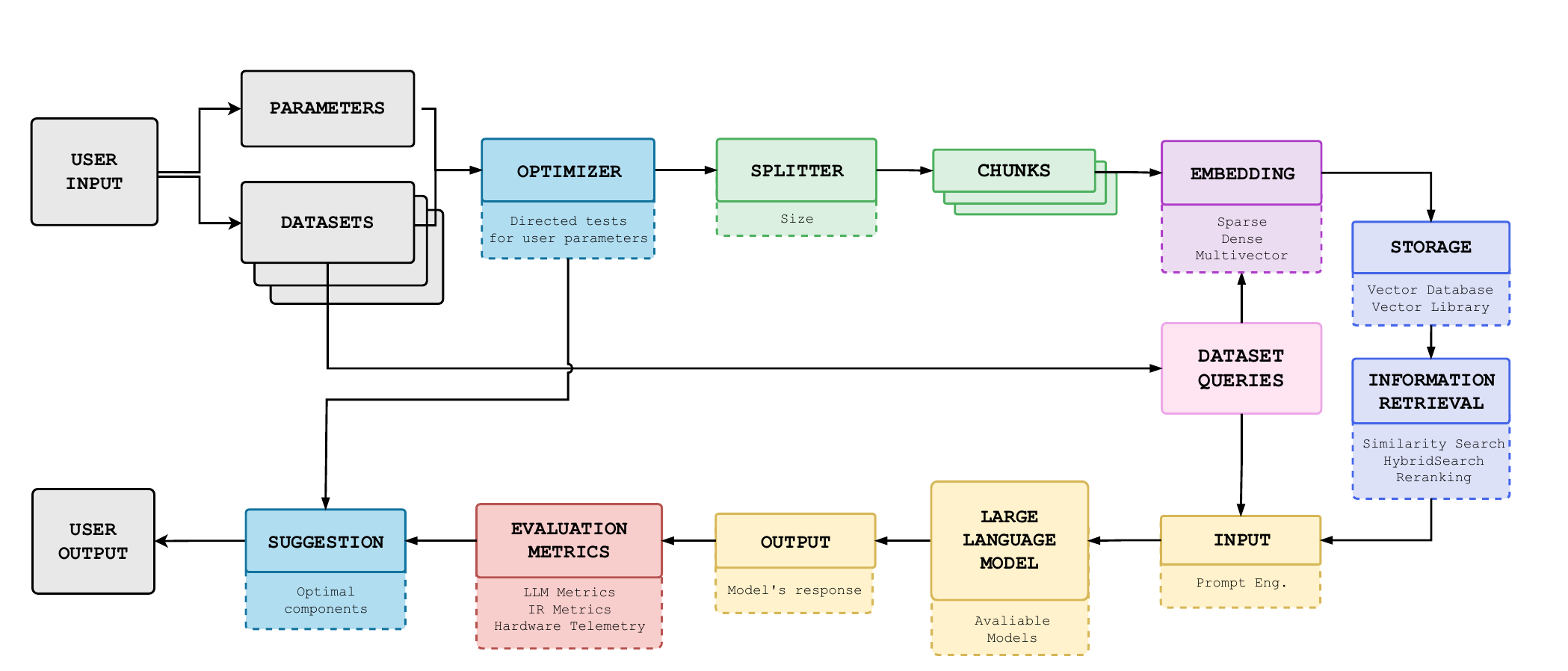}
    \caption{Overview of the RAGe framework. The user selects parameters to define what the framework must optimize, and at least one input dataset. Based on the information provided, the Optimizer component prunes options, such as models that exceed the maximum allowed memory. Once the process is done for one set of available components, the Suggestion compiles the data into visualizations to help the user decide which components are better suited to solve their task.}
    \label{fig:overview_framework}
    
\end{figure}

The main characteristic of the proposed framework is its flexibility, enabling researchers to define custom inputs, such as specific datasets, LLMs, and parameter settings, for the evaluation process.
To ensure operational continuity, the system features an automated fallback mechanism that assigns default values when explicit inputs are omitted.
Subsequently, the framework performs an optimization step by analyzing the input structure and parameters, and provides key metrics such as response time and output quality.

Internally, RAGe orchestrates an evaluation pipeline encompassing data processing, retrieval, and generation modules. 
Furthermore, the framework processes the input by passing it through a splitter module that breaks the documents into chunks. 
The system processes each chunk through an embedding model and indexes the results in VDBs. To optimize retrieval, a specialized module applies hybrid search and reranking to extract the most relevant chunks. The framework then combines these retrieved contexts directly with the user queries. After this process, the LLM module evaluates the input prompt and processes all subsequent text. 
Following this, the LLM as a judge evaluates metrics such as hallucination and faithfulness. 
Concurrently, the framework captures telemetry data to profile hardware utilization during model inference. 
The end-to-end process concludes with a comprehensive analytical report composed using the suggested RAG parameter configuration and all collected metrics for each experiment combination.

\subsection{Parameter Configuration}\label{subsec:confgis}

The parameter configuration is important for allowing the researcher to customize each module to the RAG application's context.
It accommodates standard benchmarking corpora, including NaturalQuestions, NewsQA, and TriviaQA, as well as custom, proprietary datasets.
To mitigate computational overhead, the methodology incorporates configurable sampling mechanisms to evaluate specific data subsets.

The framework systematically explores all the provided modules options, encompassing diverse vector storage solutions, retrieval algorithms, similarity metrics, reranking techniques, and text segmentation parameters such as chunk size and overlap.

To address the limitations of consumer-grade hardware, the architecture sets thresholds for maximum acceptable latency and VRAM usage.
The evaluation pipeline dynamically prunes any experimental configuration that exceeds these predefined resource limits, ensuring execution viability. Figure~\ref{fig:pruning} shows a diagram of the pruning process. A combination will be tested against the threshold values; if any exceed the thresholds, the combination is no longer tested.

The framework incorporates a weighting mechanism to prioritize metrics across three core categories: generation, retrieval, and hardware telemetry.
Researchers can assign one of four distinct priority levels for each category, ranging from no relevance to high importance. After the evaluation phase, the system utilizes these predefined preferences to recommend the optimal combination of RAG modules.

An automated recommendation engine subsequently leverages these weighted parameters to identify the optimal synergistic combination of RAG components.
The methodology decouples the experimental design from the execution environment by serializing all parameters into a standardized configuration payload.
Finally, the system features a predictive cost model that estimates the total execution time for the full combinatorial space of the experiment.
This heuristic calculates the projection based on the time to process one line with the selected VRAM and the number of different models/chosen techniques. To calculate, the system's warm-up time is disregarded.

\subsection{Optimizer}
This module initiates framework execution by checking whether any module combinations were executed in a previous run. 
If so, the optimizer compares the previous results with the input thresholds and skips executing any components that previously exceeded them. 
There are two distinct threshold values: latency and maximum VRAM. 

Latency can be either the total latency from question to answer or separated into retrieval and generation latency. 
For generation latency, after each completed execution, the mean of the combined LLM and embedding model is calculated and saved for future executions. 
The same approach applies to retrieval latency, with the primary difference being the fields used for comparison, including embedding models, storage type, search type, distance strategy, chunk size, and the number of retrieved chunks. 

\begin{figure}[H]
    \centering
    \includegraphics[width=1\linewidth]{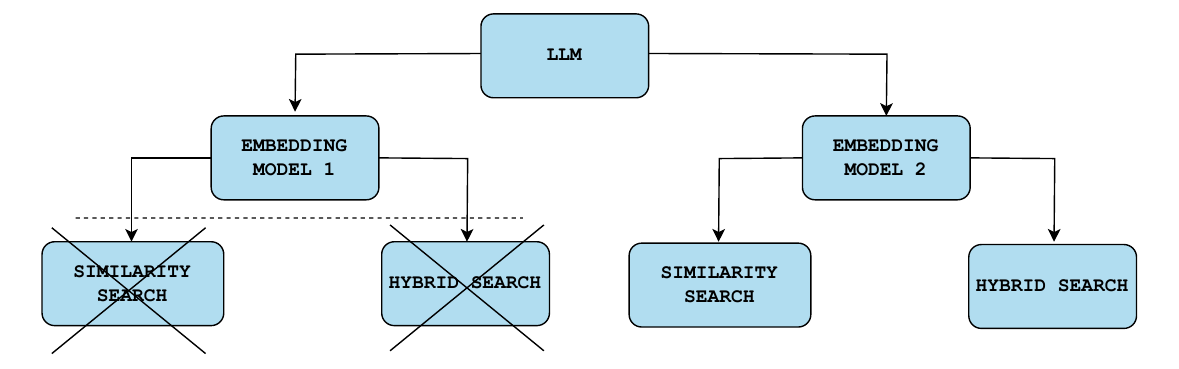}
    \caption{Pruning process for optimization of test execution. In this scenario, the combination of the LLM and Embedding 1 exceeds the threshold; therefore, it is not further tested. }
    \label{fig:pruning}
    
\end{figure}

For VRAM, the maximum value is treated as exceeding the threshold, which will interrupt execution. 
This process optimizes runtime by discarding future combinations that exceed user-defined thresholds. 
Additionally, the optimizer recognizes the initial register of a new combination and excludes it from consideration due to the time required to load models and tools.
Figure \ref{fig:pruning} shows a diagram of the pruning process.

\subsection{Suggestion}

Based on performance and results, RAGe identifies the optimal parameter combination given the user's input \ref{subsec:confgis}. User-defined weights are mapped to numerical values as follows: ``No Relevance'': 0, ``Low'': 1, ``Medium'': 3, ``High'': 5.
For each instance (k) of a given combination (j), the contribution of each metric (i), further referenced as $M_{ijk}$, is calculated by multiplying its value by its assigned weight ($w_{i}$). The weighted score, $S_{ijk}$, is determined based on the metric characteristics, as shown in Equation \ref{eq:weighted_score}. The normalization function, applied to metrics where lower values are better, ensures that all metric scores are on a comparable scale, with higher values indicating better results. min($M_{i}$) and max($M_{i}$) represent the minimum and maximum measured values of metric i across all instances.

\begin{equation}
S_{ijk} =
\begin{cases}
w_i \cdot M_{ijk} & \text{if metric } i \text{ is `high is better'} \\
w_i \cdot \left(1 - \frac{M_{ijk} - \min(M_i)}{\max(M_i) - \min(M_i)}\right) & \text{if metric } i \text{ is `low is better'}
\end{cases}
\label{eq:weighted_score}
\end{equation}

To assess the overall performance of each combination, the calculated values from all metrics and instances are aggregated into a single mean composite score $\bar{S}_j$, as shown in Equation~\ref{eq:mean_composite_score}. 

\begin{equation}
\bar{S}_j = \frac{1}{N} \sum_{i=1}^{N} S_{ijk}
\label{eq:mean_composite_score}
\end{equation}

The combination with the highest score ($J_{best}$), as determined by Equation \ref{eq:best_comb}, is the optimal solution based on the user-defined preferences.

\begin{equation}
J_{\text{best}} = \arg\max(\bar{S}_j)
\label{eq:best_comb}
\end{equation}

%% file: sections/04_toolchain.tex
\section{Toolchain}\label{sec:toolchain}

\begin{figure}[ht]
    \centering
    \includegraphics[width=0.75\linewidth]{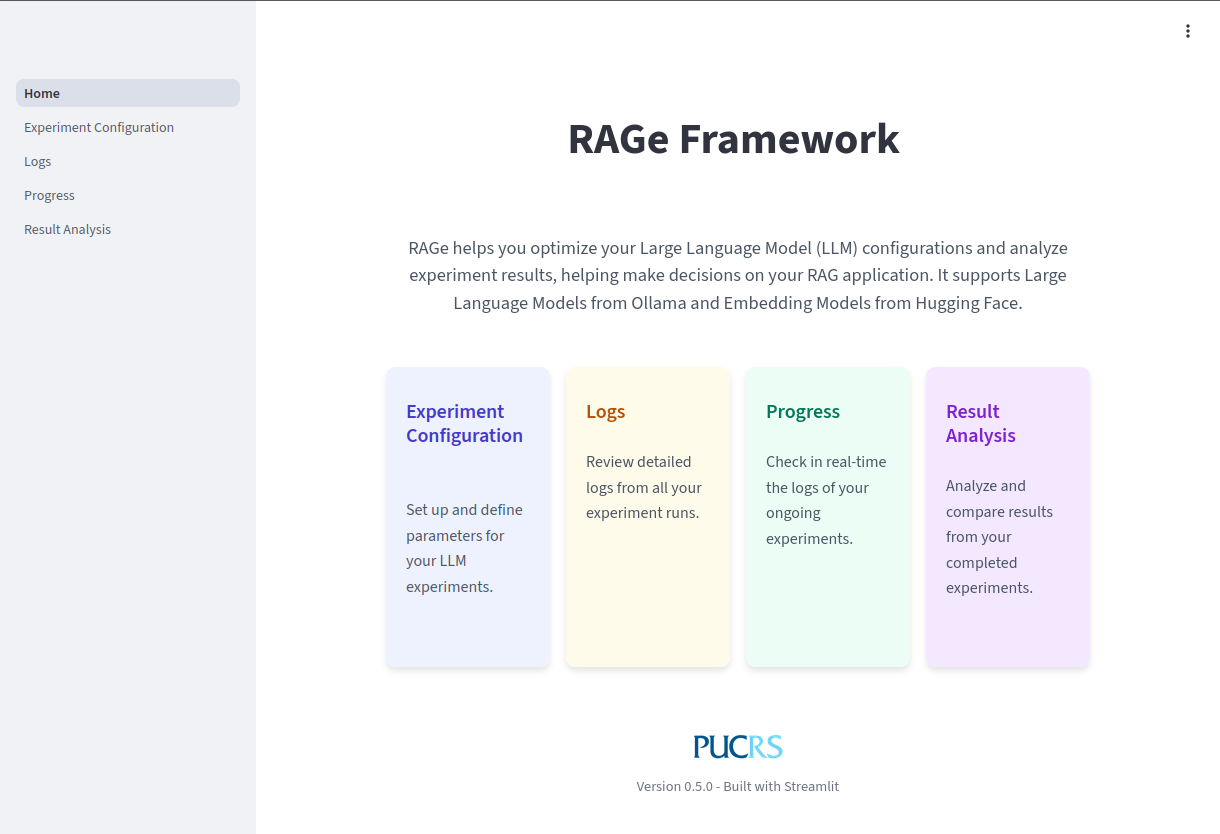}
    \caption{Homepage for toolchain. The homepage gives a brief explanation of the framework and the other pages of the UI.}
    \label{fig:homepage}
    
\end{figure}

Our toolchain features a dedicated User Interface (UI), illustrated in Figure \ref{fig:homepage}, that orchestrates the modular RAG evaluation process. Rather than hardcoding parameters, it is possible to customize each module. The experimental setup can include one or more LLMs, embedding models, and standard benchmarking corpora or custom datasets. 

Furthermore, the toolchain exposes a set of hyperparameters for the retrieval module. This enables fine-grained control over storage types, search types, similarity metrics, and reranking techniques, as well as the chunk size, chunk overlap, and the number of chunks for retrieval.

To align the evaluation with specific research objectives, the interface also supports selecting preference weights for two groups of metrics related to generation and retrieval, as well as for hardware telemetry. 

Once the experimental setup has been established, these configurations are serialized into a standardized JSON payload that serves as the definitive configuration schema for the core framework. 
Furthermore, with the file-based experiment configuration, the framework supports decoupled environments, enabling experiments to be designed on a local client but executed on remote or specialized hardware.
The subsequent subsections detail the specific components and configuration options supported by the toolchain.

\subsection{Datasets}
The toolchain's expected dataset structure must include three fields: context, question, and answer. 
Therefore, three datasets are available as default options: Natural Questions, NewsQA, and TriviaQA. 

The Natural Questions dataset has short and long answers. Since short answers are more variable and occur less frequently, we used only long answers. 
For the NewsQA dataset, each original document has multiple questions and answers; they were divided into one question per row, and the context document was copied for each instance.
TriviaQA's context documents were retrieved from the evidence directory. 

Each dataset is preprocessed and stored as a list of dictionaries containing the expected fields.
The user can select multiple default datasets from the framework and provide custom datasets; these must be JSON or CSV files.

    

\subsection{Splitter and Chunks}
Once a structured dataset is processed and formatted in JSON, the documents must be further processed for RAG. 
To do this, the toolchain uses RecursiveCharacterTextSplitter\footnote{RecursiveCharacterTextSplitter available at LangChain documentation: \url{https://python.langchain.com/api_reference/text_splitters/character/langchain_text_splitters.character.RecursiveCharacterTextSplitter.html}} to divide each document provided into smaller parts known as chunks. 

The chunk size and the number of overlapping characters in consecutive chunks can be configured in the input. 
The process of splitting text is necessary because it breaks documents into smaller and more identifiable units. 
This enables the system to retrieve and process only the most relevant portions of a document, reducing the amount of information provided to the LLM and thereby making the process more efficient.

\subsection{Embedding}

After the document splitting process, each resulting text chunk is transformed into a numeric vector representation through the application of sentence embedding models (Figure \ref{fig:embedding-process}). 
Subsequently, these embeddings are used to measure the semantic similarity between dataset queries and document chunks. 
By default, the toolchain supports any Sentence Transformer\footnote{Sentence Transformer Framework available at: \url{https://huggingface.co/sentence-transformers}} embedding model, such as \textit{jina-embeddings-v3} and \textit{all-MiniLM-L6-v2}.
These sentence embeddings can be stored in a vector database or in a vector library, depending on input parameters. 
Vector representations are generated incrementally as chunks are inserted into the chosen storage system.

\begin{figure}[tb]
    \centering
    \includegraphics[width=0.75\linewidth]{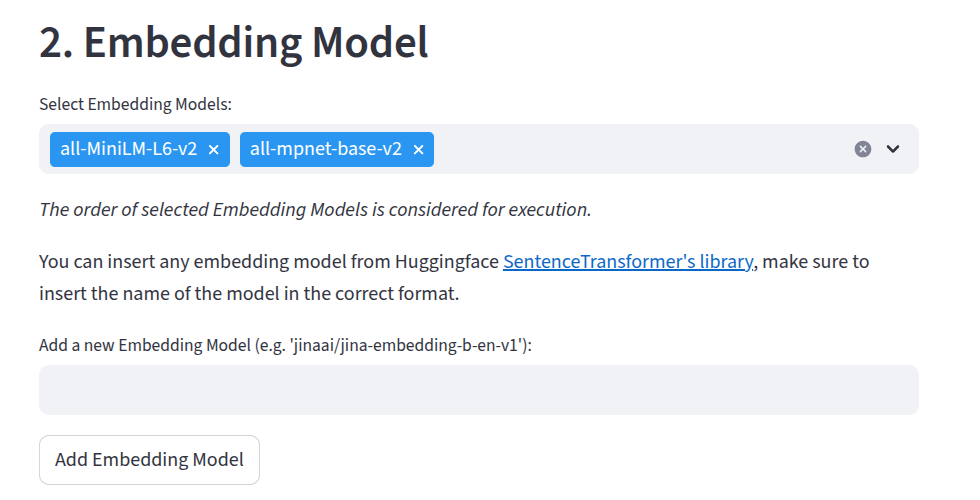}
    \caption{In the experiment configuration page, as shown in Section~\ref{subsec:confgis} the user selects an LLM. They can also select embedding models from Hugging Face. If the models are not yet installed, they are automatically downloaded from the sentence-transformers library.}
    \label{fig:embedding}
\end{figure}

\subsection{Storage} 
\label{subsec: mstorage}
To support the retrieval of embedded documents, the toolchain offers two storage options: a vector database and a vector library. 
Vector databases are used for information retrieval, managing large-scale document collections, and supporting full CRUD (Create, Read, Update, and Delete) operations. 
In contrast, vector libraries store numerical vectors in memory, prioritizing fast retrieval performance at the cost of limited functionality, as the data is volatile and does not support all CRUD operations. 
By default, the toolchain integrates PGVector\footnote{PGVector implemented via LangChain available at: \url{https://python.langchain.com/docs/integrations/vectorstores/pgvector/}} as the vector database and FAISS\footnote{FAISS implemented via LangChain available at: \url{https://python.langchain.com/docs/integrations/vectorstores/faiss/}} as the vector library, both using the LangChain implementation, and can be selected via input parameters.

\subsection{Information Retrieval} 
To retrieve documents from the vector storage, two main retrieval strategies were employed: similarity search and hybrid search. These approaches enable the system to identify and retrieve the most relevant document chunks given a natural language query.

The implemented similarity search method supports three distance metrics: Cosine Similarity, Euclidean distance, and Inner Product. These metrics measure the semantic distance between the embedded query and document chunks to retrieve the most relevant content. For the hybrid search approach, we incorporate into the similarity search method a sparse retrieval step using PostgreSQL's tsvector\footnote{tsvector, available at: \url{https://www.postgresql.org/docs/current/datatype-textsearch.html}} to identify the top-$k$ relevant chunks based on lexical similarity. These results are then refined in a reranking phase, in which the selected chunks are sorted by their retrieval scores. The value of $k$, along with the retrieval approach and similarity distance metrics, is defined in the input parameters.

To further enhance retrieval quality, a reranking phase is introduced. In our implementation, each document is ranked and sorted based on \textit{jina-reranker-v2-base-multilingual}\footnote{jina-reranker-v2-base-multilingual, available at: \url{https://huggingface.co/jinaai/jina-reranker-v2-base-multilingual}}, a pretrained model. After information retrieval and reranking, the content is passed to the language model for generation.

\subsection{Queries, Prompt and LLM}
We build the prompt by integrating the input query with the retrieved information. This prompt is then sent to Ollama\footnote{Ollama, available at: \url{https://ollama.com/}} to generate a response. Ollama supports multiple LLMs and simplifies the generation process. After the selected model generates a response, it is evaluated using the evaluation metrics.
Using a zero-shot technique~\cite{mao2025prompts}, we used the following prompt:\\

\begin{verbatim}
    You are a professional assistant for answering questions.
    Your task is to summarize the topic using 
    information from a given CONTEXT
    You will receive CONTEXT information 
    and a user QUERY. Keep your answer grounded in 
    the facts of the CONTEXT.
    You have to be as concise as possible, 
    while still answering the user's query.
    If the CONTEXT doesn’t contain the facts to 
    answer the QUERY, return No relevant information provided.
    Answer the users QUERY using the CONTEXT. 
    CONTEXT information will be provided in the next sentence. 
    {similar_chunks}
    Here is the user query: {query}
    
\end{verbatim}

\subsection{Evaluation Metrics}
To evaluate each generated text, the toolchain uses DeepEval\footnote{DeepEval, available at: \url{https://github.com/confident-ai/deepeval}}, which offers a wide variety of metrics to evaluate RAG systems. These metrics encompass both text generation and context retrieval; these are further elaborated in Section \ref{subsec:metrics}. In addition, we also collect system telemetry data (e.g., hardware usage) using psutil\footnote{psutil, available at: \url{https://github.com/giampaolo/psutil}} and pynvml\footnote{pynvml, available at: \url{https://pypi.org/project/pynvml/}}. psutil retrieves information on running processes and system utilization, while pynvml provides information on GPU management and monitoring functions.

%% file: sections/05_results.tex
\subsection{User Interface}\label{sec:results}

Toolchain UI is structured into five pages: Home, Experiment Configuration, Logs, Progress, and Result Analysis. The application is built using the \textit{Streamlit} library.
On the home page (Figure~\ref{fig:homepage}), it is possible to explore the pages and find a brief explanation of the toolchain.

    


During execution, the toolchain features a dynamic tracking system that allows researchers to monitor the real-time performance of each module combination. 
The active session interface reflects the assigned evaluation weights and offers an early preview of the visual analytics. 

\begin{figure}[H]
    \centering
    \includegraphics[width=1\linewidth]{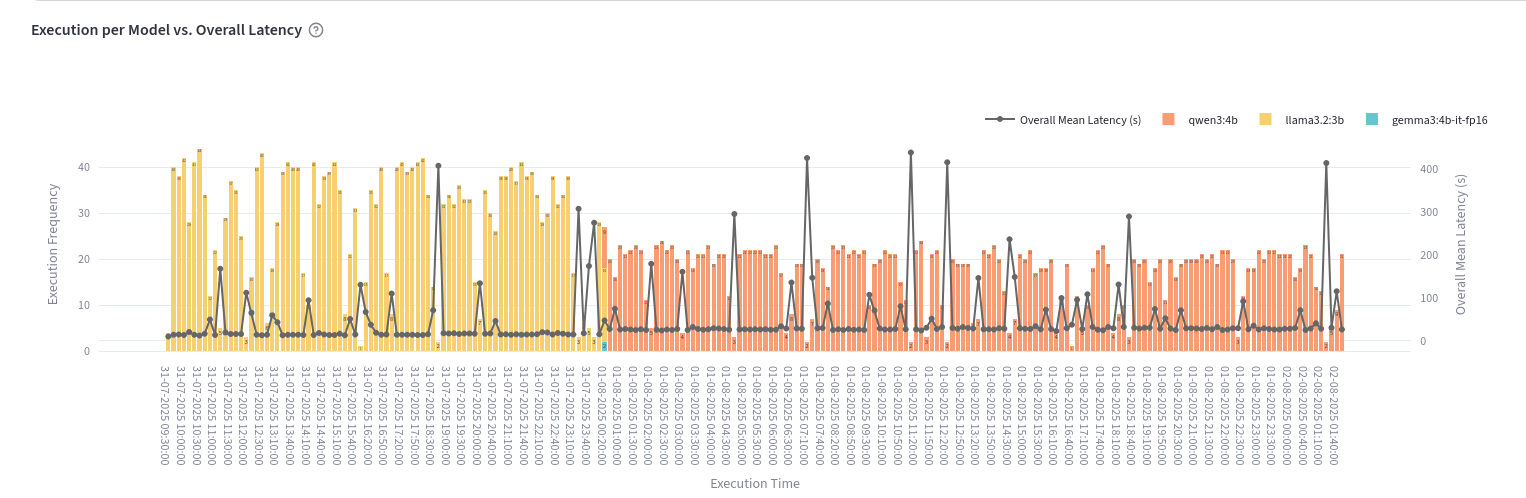}
    \caption{In result analysis section, multiple visualizations are created from the results file, such as number of traces in a given time. By showing how many traces are presented in a specific period of time and showing models in different colors, we can easily check if a model generates responses faster than another.}
    \label{fig:result_analysis}
\end{figure}

\begin{figure}[H]
    \centering
    \includegraphics[width=0.8\linewidth]{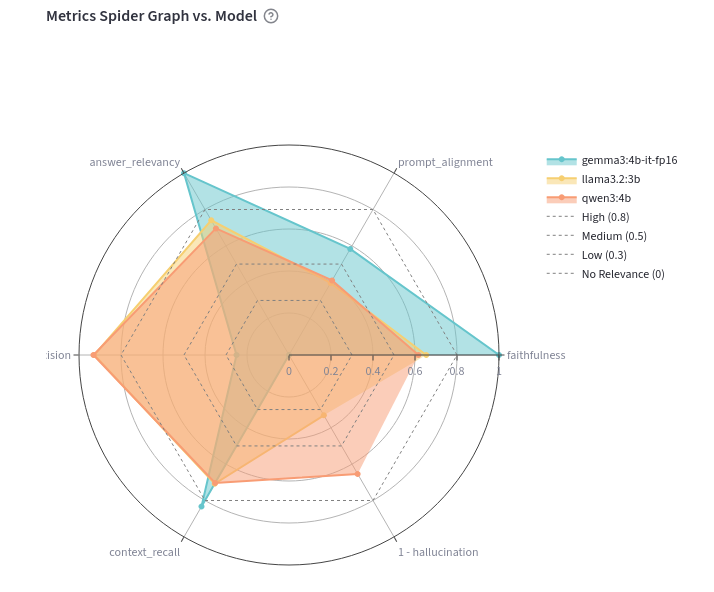}
    \caption{The spider graph shows the difference between LLMs. To address the difficulty of radial graphs, the toolchain also displays a continuous line for high, medium, and low values, providing a baseline for comparing models.}
    \label{fig:spider_graph}
\end{figure}

Post-execution, the toolchain generates a graphical analysis to facilitate performance interpretation. This includes chronological trace graphs to compare model latency, as shown in Figure~\ref{fig:result_analysis}, and radar (spider) charts, to visualize performance across various LLM-as-a-judge metrics, as shown in Figure~\ref{fig:spider_graph}. For visual coherence on these charts, inversely proportional metrics are adjusted. The hallucination metric is normalized and plotted as the inverse score (1 - hallucination).

\begin{figure}[H]
    \centering
    \includegraphics[width=1\linewidth]{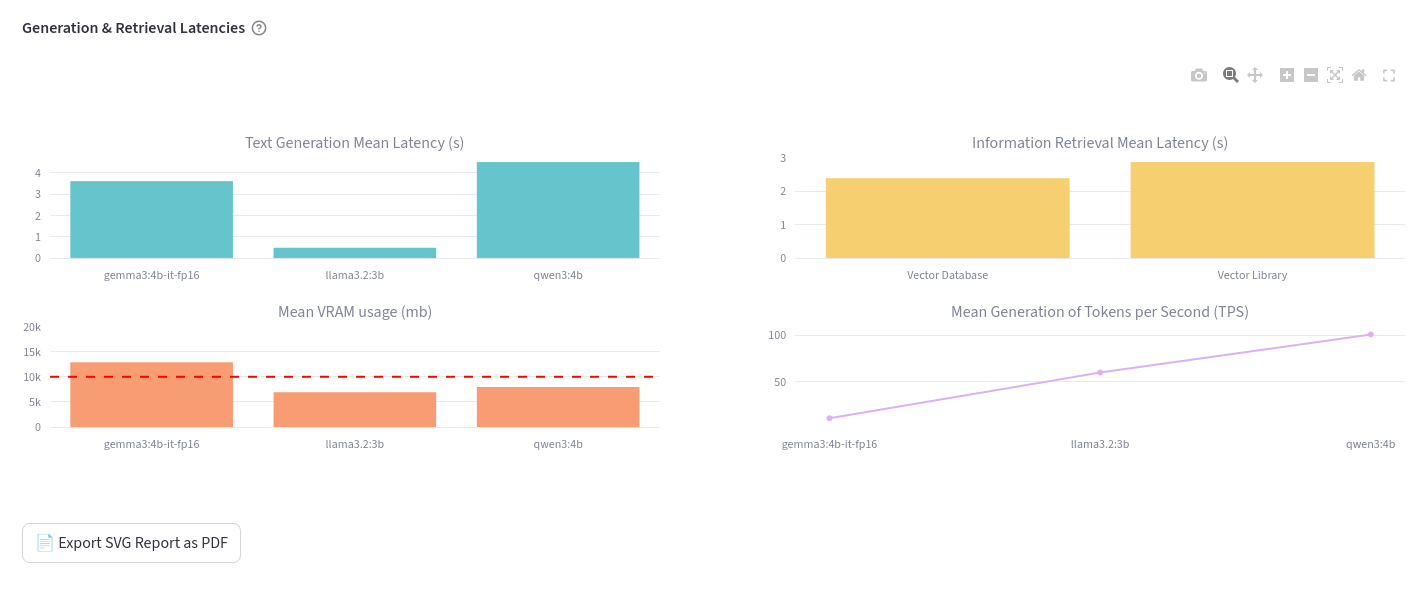}
    \caption{Generation and retrieval latencies visualization from each model and each storage type.}
    \label{fig:last_vis_ra}
\end{figure}

The last visualization in Figure~\ref{fig:last_vis_ra} shows the generation and retrieval latencies from each model and each storage type. The mean VRAM usage is displayed, and the maximum VRAM threshold set in the configuration is shown as a dotted line. The mean token generation rate per second for each model enables deeper analysis of generation, since a model might take longer to generate an answer but produce longer answers.

\begin{figure}[H]
    \centering
    \includegraphics[width=1\linewidth]{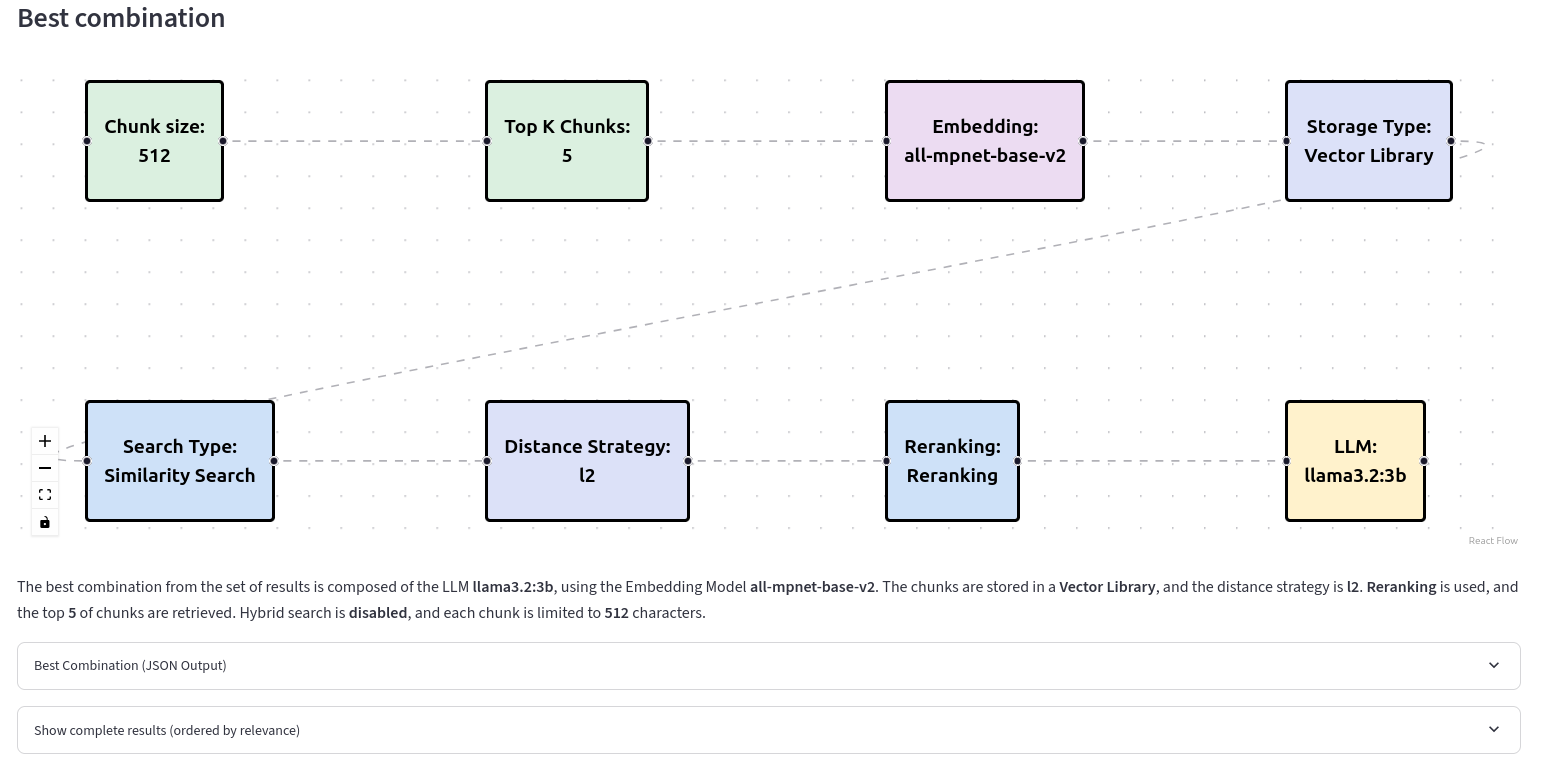}
    \caption{Best combination from a given set of results. Displays a dynamic visualization using Streamlit Flow, accompanied by a text stating which elements yield the best results. It is possible to pan and zoom on the different elements of the recommended RAG system.}
    \label{fig:best_combination_rel}
\end{figure}

Following an experimental session, the toolchain post-execution module synthesizes the evaluation metrics to automatically construct an optimal RAG configuration tailored to the predefined scenario parameters, as illustrated in Figure~\ref{fig:best_combination_rel}. 
To facilitate seamless deployment and iterative testing, the toolchain dynamically generates a new, ready-to-use configuration file based on these data-driven recommendations.

The architectural requirements for RAG pipelines are inherently context-dependent. While certain applications may benefit from greater generative flexibility, others are constrained by computational resources. 
The main advantage of using the RAGe framework is that it abstracts the process of testing each possible configuration across the available services and models. This can be hundreds of combinatorial possibilities. 


%% file: sections/06_rw.tex
\section{Related Work}\label{sec:rw}

In this section, we present the most recent related work in the literature.
We consider papers published in conferences and journals, as well as the grey literature, which includes preprints available on arXiv.
We focus on modular RAG frameworks and found seven different approaches: Modular RAG, RAGLAB, Ragnarök, XRAG, Faster-Cheaper-Better, FlashRAG, and OptiRAG.
All proposed approaches are from 2024 to 2025.

\subsection{Modular RAG}

The Modular RAG proposed by \citet{modular_rag} provides a structured and flexible framework for designing RAG systems by decomposing them into clear, functional, Lego-like modules.  The framework comprises six modules that can be configured independently to meet user needs. Each module employs different techniques, models, and metrics.

According to the proposal, the pre-retrieval contains query transformation, expansion, and construction as options. The constructions can transform natural language into structured query formats, such as Cypher or SQL.
The hybrid retrieval stage then integrates fine-tuning capabilities, multi-source routing, and dynamic selection strategies.
Subsequently, the post-retrieval module applies reranking, contextual compression, and an LLM-guided selection process to isolate the most highly relevant chunks.
Finally, for generation, there is a generator with fine-tuning and a verification step that verifies the hallucination contains private data and whether it answers the question.

In parallel with the retrieval and generation pipelines, an orchestration module dynamically manages the system's execution flow. 
It features a semantic routing system in which an evaluator judges the query to determine whether the response should stem from the retrieved context or the model's inherent knowledge base.
Furthermore, a scheduling component continuously monitors the pipeline to determine whether to trigger subsequent retrieval executions. 
Finally, a knowledge-guided module utilizes structured reasoning paths to navigate and resolve complex, multi-step queries.

\subsection{RAGLAB}

~\citet{raglab} proposes the RAGLAB framework, a modular, research-oriented framework designed to support low-level control and experimentation to address the problems faced by researchers in the area.
Regarding the authors' algorithmic solutions, the focus is on simplifying algorithm development through a plug-and-play architecture. It provides modular components that users can manipulate and combine to build custom RAG pipelines.
Additionally, users can choose to interact with the framework via its UI or integrate it into their code via a library.

RAGLAB shows notable features as generator module, which is a feature that integrates huggingface transformers and VLLM, allowing efficient inference by supporting techniques such as Low-Rank Adaptation (LoRA)~\cite{hu2022lora} and Quantized LoRA (QLoRA)~\cite{dettmers2023qlora}.
Furthermore, a module that is important for architecture is the instruction lab, which consists of three key prompt types - System Instruction, Task Instruction, and Algorithm Instruction - making it versatile for users to choose specific prompts to specific domains or use cases.

The trainer module is also implemented, using Accelerate\footnote{Trainer module available at https://github.com/huggingface/accelerate} and Deepspeed \footnote{Training optimizer available at https://github.com/deepspeedai/DeepSpeed} libraries to enable scalable and efficient fine-tuning. 
Referring to datasets, metrics, and algorithms, RAGLAB implements five types of datasets: OpenQA, Multi-HopQA, Multiple-Choice, Fact Verification, and Long-Form QA.
It incorporates both classic and advanced evaluation metrics, including accuracy, exact match, F1 score, Factscore, and ALCE.
In addition, the framework has six published algorithms: Naive RAG, RRR, ITERRETGEN, Self-ASK, Active RAG, and SelfRAG.





\subsection{Ragnarök}

\citet{pradeep_ragnarok_2025} developed the Ragnarök framework. The authors propose a framework consisting of two modules: retrieval and augmented generation. The framework's main contribution is its reusability throughout the end-to-end pipeline.  
The retrieval module's standard implementation utilizes lexical retrieval models, specifically BM25, while RankZephyr is used for reranking. 
The retrieval process retrieves the top 100 most similar chunks and applies reranking to obtain the top 20. In sequence, the top 20 are used as input to the LLM to generate the answer in the Augmented Generation module. The generated answer contains each original document as a reference for each sentence. Command R+ and GPT4o LLM models are used.

One of the key contributions is a user interface (UI) that enables users to compare answers from both models based on the same query. The UI has two modes: blind and non-blind, where users can see the models being used.
With a blind approach, bias can be avoided, allowing users to choose the best retriever, reranker, and LLM model based on the answer.
The tool is open-source and can be found on the official GitHub repository~\footnote{Ragnarök official GitHub repository: \url{ https://github.com/castorini/ragnarok}}.
In addition to the UI, the framework can be used with the REST API. 

\subsection{XRAG}
\citet{xrag}\footnote{As of the date July 28, 2025, the paper introducing the XRAG framework is available only as a preprint. XRAG official GitHub repository \url{https://github.com/DocAILab/XRAG}
} proposes the evaluation in four stages: pre-retrieval, retrieval, post-retrieval, and generation.
The authors utilized datasets to evaluate HotpotQA, DropQA, and NaturalQA.
XRAG is available for use via the UI and the API.
On pre-retrieval, there are implementations of Step-back Prompting (SBPT) and Hypothetical Document Embedding (HyDE). There are two approaches to retrieval: basic, with BGE-Large and Jina-Large, and advanced, which integrates LlamaIndex. With LlamaIndex, the following approaches are used: Reciprocal Rerank Fusion Retriever (RRFusion) and SentenceWindow Retriever (StParser).
The generation offers integration with OpenAI, DeepSeek R1-7B, HuggingFace, and Ollama API. It is necessary to configure the OpenAI and HuggingFace tokens.
Post-retrieval refinement, summarization, or compaction ensures contextual clarity. Supports BGE reranker, Jina-Reranker V2 and ColBERTv2.
XRAG does not support hardware telemetry, and does not have a graphical report result. 
Conventional Retrieval Evaluation (ConR) for retrieval unit matching, Conventional Generation Evaluation (ConG) for generative token-matching based generation tests, and Cognitive LLM Evaluation (CogL) for semantic understanding based generation tests

\subsection{Faster, Cheaper, Better}
~\citet{fast-cheaper-better} propose a multi-objective optimization approach for end-to-end RAG systems in Faster, Cheaper, Better: Multi-Objective Hyperparameter Optimization for LLM and RAG Systems. 
Their method performs optimization over the entire pipeline, including LLM selection (e.g., GPT-4o, GPT-4o-mini, Llama-3.2-3B, and Llama-3.1-8B), embedding models (text-embedding-3-large and text-embedding-3-small), chunking strategies, reranking thresholds, and generation parameters such as temperature. 
The article also introduces two datasets—FinancialQA and MedicalQA—designed to reflect real-world industry RAG scenarios.

The authors formulate the RAG configuration as a multi-objective optimization (MOO) problem and aim to identify Pareto-optimal configurations that balance four competing objectives: cost, latency, safety (hallucination risk), and alignment (helpfulness). 
In this setting, a configuration is considered Pareto-optimal if no other configuration improves one objective. 
To approximate the Pareto frontier, they employ Bayesian Optimization (BO) with the qLogNEHVI acquisition function. 

Their results demonstrate that Bayesian optimization significantly outperforms the baseline, obtaining superior Pareto fronts across two benchmark tasks.
This work represents a systematic application of multi-objective Bayesian optimization for discovering trade-offs in RAG pipelines.
However, selecting configurations from the Pareto frontier remains a challenge in deployment scenarios that require dynamic trade-offs among objectives.

\subsection{FlashRAG}
\citet{flashrag} proposes the FlashRAG framework, which has five core modules and sixteen diverse RAG subcomponents that can be integrated independently or combined into pipelines. 
Additionally, it offers nine standardized RAG processes, a UI, and auxiliary scripts for corpus construction, building retrieval indexes, and preparing retrieval results.

The structure comprises five elements: judger, retrieval, reranker, refiner, and generator.
The Judger is used to determine if a query needs retrieval.
The retrieval implementations are based on sparse models using BM25, as well as dense models employing BERT-based and T5-based approaches. For vector operations, the FAISS library is used. The dataset management is through HuggingFace.
On reranking, BGE and JINA are employed as reranker models.
The refiner is based on extractive, abstractive, and perplexity-based approaches. 
The final component in the RAG process is the generator. It integrates advanced LLM acceleration libraries like vLLM and FastChat. It also provides the native interface of the Transformers library.
Default models used for benchmarking include LLaMA-3-8B-instruct and Qwen-1.5-14B.
It also includes encoder-decoder models, such as Flan-T5, and uses fusion-in-decoder techniques.

The pipeline module implements four different process flows: sequential, branching, conditional, and loop. 
FlashRAG implements 38 benchmark datasets, standardized in JSONL format, covering different tasks.
The evaluation metrics are based on retrieval, generation, and the number of tokens to estimate costs.
The retrieval-aspect metrics are Recall@k, precision@k, F1@k, and mean average precision (MAP).
For generations, the metrics include Token-Level F1 score, exact match, accuracy, BLEU, and ROUGE-L.

\subsection{OptiRAG}
~\citet{OptiRAG} proposes an optimization framework for RAG pipelines, closest to our work, modeling of the system behavior across combinations of hyperparameters. 
Their approach adopts a modular composition of components — including LLMs, embedding models, chunking strategies, and retrieval parameters — and employs regression-based performance modeling combined with search to identify configurations that optimize latency and cost efficiency.

OptiRAG formulates the problem using Pareto-based multi-objective optimization (also explored in ~\citet{fast-cheaper-better}), aiming to identify configurations that balance competing system-level objectives. 
This formulation is particularly suitable for deployment-oriented scenarios, as it emphasizes computational efficiency while reducing the need for exhaustive exploration of the combinatorial search space.

For embedding selection, the authors use the Massive Text Embedding Benchmark (MTEB) ~\citet{mmteb} to filter candidate models according to reported benchmark performance within a given parameter range. 
At the retrieval level, the framework evaluates VDBs such as FAISS and ChromaDB, modeling their hyperparameters and impact on cost, latency. 
For LLMs selection, OptiRAG based on CAR-LLM \cite{krishnan2024car}, which recommends the model and hardware based on workload characteristics such as input length, output length, and query volume.

\subsection{Discussion}\label{sec:discussion}

These frameworks and approaches presented are generally structured around similar core modules, such as data ingestion and RAG, yet each adopts distinct approaches to model, dataset usage, and evaluation. 
Despite these differences, the main objective across all frameworks is to provide effective RAG solutions with robust features and a clear, understandable architecture that supports practical implementation and experimentation.
In this section, we compare the related work with our approach.

\begin{table}[ht]
\caption{The table provides a comparison of RAG Libraries and Frameworks. It evaluates the framework based on five categories: datasets used, model support platform, inclusion of a telemetry module, and optimization suggestions.}\label{tab:rw_table}
\setlength{\tabcolsep}{0.1em} 
{\renewcommand{\arraystretch}{1.2}
\begin{tabular}{cccccc}
\hline
\multicolumn{1}{c}{\textbf{Framework}} &
  \textbf{Datasets} &
  \textbf{\begin{tabular}[c]{@{}c@{}}Models\\ Support\\ Plataform\end{tabular}} &
  \textbf{\begin{tabular}[c]{@{}c@{}}Telemetry\\ Module\end{tabular}} &
  \textbf{\begin{tabular}[c]{@{}c@{}}Optimization\\ Suggestion\end{tabular}} \\ \hline
RAGe (ours, 2026) &
  \begin{tabular}[c]{@{}c@{}}\\Natural Questions,\\ NewsQA,\\ and TriviaQA\end{tabular} &
  \begin{tabular}[c]{@{}c@{}}\\Hugging Face,\\ Ollama\end{tabular} &
  \ding{51} &
  \ding{51} \\
\begin{tabular}[l]{@{}c@{}}OptiRAG\\(\citeauthor{OptiRAG}, \citeyear{OptiRAG})\end{tabular} &
  No Mention &
  No Mention &
  \ding{51} &
  \ding{51} \\ 
\begin{tabular}[c]{@{}c@{}}FlashRAG\\(\citeauthor{flashrag}, \citeyear{flashrag})\end{tabular} &
  \begin{tabular}[c]{@{}c@{}}\\Natural Questions,\\ TriviaQA,\\ PorQA + 35 integrated\\ with Hugging Face\end{tabular} &
  Hugging Face &
  \ding{55} &
  \ding{55} \\
\begin{tabular}[c]{@{}c@{}}Faster, Cheaper, Better\\(\citeauthor{fast-cheaper-better}, \citeyear{fast-cheaper-better})\end{tabular} &
  \begin{tabular}[c]{@{}c@{}}\\FinancialQA and\\ MedicalQA\end{tabular} &
  No Mention &
  \ding{55} &
  \ding{51} \\
\begin{tabular}[c]{@{}c@{}}XRAG\\(\citeauthor{xrag}, \citeyear{xrag})\end{tabular} &
  \begin{tabular}[c]{@{}c@{}}\\Natural Questions,\\ TriviaQA,\\ WebQA and HotpotQA\end{tabular} &
  No Mention &
  \ding{55} &
  \ding{55} \\
\begin{tabular}[c]{@{}c@{}}Ragnarök\\(\citeauthor{pradeep_ragnarok_2025}, \citeyear{pradeep_ragnarok_2025})\end{tabular} &
  \begin{tabular}[c]{@{}c@{}}\\ClueWeb22 and\\ MS MARCO V2\end{tabular} &
  No Mention &
  \ding{55} &
  \ding{55} \\ 
\begin{tabular}[c]{@{}c@{}}RAGLAB\\(\citeauthor{raglab}, \citeyear{raglab})\end{tabular} &
  \begin{tabular}[c]{@{}c@{}}\\PopQA,\\ TriviaQA,\\ HopPopQA,\\ WikiMultiHop, ARC,\\ MMLU, PubHealth,\\ StrategyQA,\\ Factscore and ASQA\end{tabular} &
  Hugging Face &
  \ding{55} &
  \ding{55} \\
\begin{tabular}[c]{@{}c@{}}Modular RAG\\(\citeauthor{modular_rag}, \citeyear{modular_rag})\end{tabular} &
  No Mention &
  No Mention &
  \ding{55} &
  \ding{55} \\ \hline
\end{tabular}%
}

\end{table}

Table~\ref{tab:rw_table} presents a comparative analysis of different modules and components of RAG, highlighting aspects relevant to the construction and use of these frameworks. 
Regarding the datasets, FlashRAG \cite{flashrag} and RAGLAB \cite{raglab} incorporate a wide variety of QA datasets. This diversity improves model generalization. 
In contrast, frameworks such as RAGe (our framework), XRAG \cite{xrag}, and Ragnarök \cite{pradeep_ragnarok_2025} rely on fewer datasets, thereby limiting the scope of experimentation.

When analyzing model support, RAGe, FlashRAG, and RAGLAB highlight their direct integration with Hugging Face\footnote{Site available at: https://huggingface.co/}, RAGe being the only framework to also mention compatibility with Ollama\footnote{Site available at:  https://ollama.com/}, offering a wide possibility of testing different models.
However, many frameworks remain limited to textual processing only, as seen in RAGLAB and in ours, which may be the next step for further studies in this field.

An important consideration for end users is the availability of telemetry resources and optimization recommendations, which are essential for monitoring RAG systems in production environments. 
Among the frameworks, only RAGe and OptiRAG~\cite{OptiRAG} implement both telemetry and optimization features, providing performance metrics configuration suggestions. 
In contrast, Faster, Cheaper, Better~\cite{fast-cheaper-better} focuses exclusively on optimization through multi-objective search, without incorporating a telemetry module for runtime monitoring.
Regarding these system-level capabilities, our RAGe framework employs a hierarchical, tree-based pruning strategy that reduces the search space across components prior to full pipeline evaluation.
Furthermore, it integrates a UI designed for internal deployment scenarios, facilitating reproducibility, experimentation, and usability.

In general, the analysis of Table \ref{tab:rw_table} reveals two main topics in the RAG frameworks: (i) inclusion of datasets for model training and evaluation, and (ii) the need for monitoring and optimization tools, still underexplored in most works.

%% file: sections/07_conclusion.tex
\section{Conclusion}\label{sec:conclusion}

Building RAG applications is a complex, time-consuming task. Each stage in RAG offers a wide variety of options, and each option has many customization parameters. 
Focusing on this, recent approaches propose modular frameworks for evaluating options and parameter customizations. 
However, these approaches fall short of optimizing the effort required to run all possible combinations in the RAG pipeline. 

Based on this, we proposed RAGe, a new modular RAG framework to evaluate the performance of RAG pipelines in resource-constrained environments. Our main contributions are divided into four: (i) hardware telemetry and qualitative generation metrics, (ii) component optimization to reduce experimental space exploration, (iii) configuration suggestion based on the trade-off multidimensional metrics, (iv) isolating the configuration experiments from the execution.

Because 'LLM-as-a-judge' is used for qualitative assessments, metrics like faithfulness and hallucination rates are inherently susceptible to the evaluator model's intrinsic biases. Also, our implementation is based on linear retrieval pipelines. Additionally, as future work, we intend to extend the RAGe architecture to support Multimodal Retrieval-Augmented Generation (mRAG). As multimodal large language models (MLLMs) become increasingly prevalent, adapting RAGe to evaluate the retrieval and generation of cross-modal data, will be crucial for benchmarking complex applications on resource-constrained devices.

